\documentclass[10pt,twocolumn]{IEEEtran}
\usepackage{multicol}

\usepackage{color}
\usepackage{soul}
\usepackage{amsmath}
\usepackage{graphics}
\usepackage{setspace}
\usepackage{cite}
\usepackage{latexsym}
\usepackage{float}
\usepackage{epsfig}
\usepackage{multirow}
\usepackage[table,xcdraw]{xcolor}
\usepackage{cite,cases,url}
\usepackage{amssymb}
\usepackage{graphicx}
\usepackage{epstopdf}
\usepackage{balance}
\usepackage{subcaption}
\usepackage{enumerate}
\usepackage{array}
\usepackage{algorithmic}
\usepackage{algorithm}
\usepackage{enumitem}
\usepackage[normalem]{ulem}
\usepackage{longtable}
\usepackage{ragged2e}
\usepackage{booktabs}
\usepackage{tabularx}

\useunder{\uline}{\ul}{}

\newcolumntype{L}{>{\centering\arraybackslash}m{5cm}}
\newcolumntype{K}{>{\centering\arraybackslash}m{6cm}}
\newcolumntype{P}{>{\centering\arraybackslash}m{2.3cm}}
\newcolumntype{M}{>{\raggedright\arraybackslash}m{2cm}}
\newcolumntype{N}{>{\raggedright\arraybackslash}m{2.5cm}}

\begin{document}

\title{%Next Generation Cellular Networks
Soft Tester UE: A Novel Approach for Open RAN Security Testing }
%\hl{\hl{Myths and Facts of} }
\author{
\IEEEauthorblockN{Joshua Moore, Aly S. Abdalla, Charles Ueltschey and Vuk Marojevic \\
} 
\normalsize\IEEEauthorblockA{Department of Electrical and Computer Engineering,  Mississippi State University, USA} \\
\normalsize\IEEEauthorblockA{\{jjm702,asa298,cmu32,vm602\}@msstate.edu}
}
\maketitle
%TC:ignore 
\begin{abstract}
With the rise of 5G and open radio access networks (O-RAN), there is a growing demand for customizable experimental platforms dedicated to security testing, as existing testbeds do not prioritize this area. Traditional, hardware-dependent testing methods pose challenges for smaller companies and research institutions. The growing wireless threat landscape highlights the critical need for proactive security testing, as 5G and O-RAN deployments are appealing targets for cybercriminals. To address these challenges, this article introduces the Soft Tester UE (soft T-UE), a software-defined test equipment designed to evaluate the security of 5G and O-RAN deployments via the Uu air interface between the user equipment (UE) and the network. The outcome is to deliver a free, open-source, and expandable test instrument to address the need for both standardized and customizable automated security testing. By extending beyond traditional security metrics, the soft T-UE promotes the development of new security measures and enhances the capability to anticipate and mitigate potential security breaches. 
The tool's automated testing capabilities are demonstrated through a scenario where the Radio Access Network (RAN) under test is evaluated when it receives fuzzed data when initiating a connection with an UE.
\end{abstract}
%TC:endignore
\IEEEpeerreviewmaketitle

\section{Introduction}
\label{sec:intro}
The rapid evolution of 5G and open radio access networks (O-RAN) necessitates the development of robust and flexible security testing mechanisms. 
Despite the increasing availability of 5G testbeds and research tools, 
customizable experimental platforms that can 
satisfy diverse requirements are still needed. Traditional testing approaches 
characterized by high costs associated with proprietary hardware and closed-source frameworks, present significant barriers towards 
achieving comprehensive security testing. Moreover, the dynamic wireless threat landscape, marked by 
sophisticated cyber threats targeting 5G and O-RAN deployments, underscores the %critical 
necessity for proactive and %thorough 
advanced security testing techniques. %protocols.

Open RAN architectures, with their disaggregated structure and open interfaces facilitating multi-vendor interoperability, introduce numerous vulnerabilities that make them prime targets for %cyber 
security attacks~\cite{10467183}. %, as extensively documented in recent literature  
These vulnerabilities include potential exploits %at various points within the network 
at different architecture components, ranging from the radio access network to the core network. %elements. 
To address these challenges, %requires 
advanced security testing tools capable of assessing authentication mechanisms, encryption protocols, %communication security, 
and the resilience of disaggregated network components under %simulated 
various attack scenarios are required.

%Existing 
Currently, security testing tools are limited within the literature. For example, tools like the A1 interface testing tool in \cite{10.1145/3643833.3656118} and those focused on Denial of Service (DoS) attacks on fronthaul interfaces in \cite{feliana2024evaluation} and \cite{dosattacker} are tailored to specific functionalities. However, much of the existing literature focused on testing operates in an ad-hoc manner, focusing on achieving stand-alone testing objectives rather than conducting comprehensive security assessments across the entire O-RAN ecosystem. An AI testing framework, as demonstrated in \cite{oaict}, exemplifies a comprehensive approach to testing but is focused on evaluating AI models within O-RAN deployments and does not consider security tests. In contrast, the framework proposed in \cite{janus} offers extensive customization of telemetry and code execution within O-RAN architectural components. However, it lacks built-in security testing capabilities and necessitates architectural modifications.

To address these
challenges, 
this article introduces the Soft Tester UE (soft T-UE), a software-defined testing solution designed  
for evaluating the security posture of 5G and O-RAN deployments. This open-source tool, compatible with commercial-off-the-shelf (COTS) software radio hardware, empowers users to conduct  
standardized and customized security tests. The soft T-UE aims to serve as an expandable testing platform equipped with sample test cases, comprehensive documentation, automated CI/CD pipelines, performance benchmarks, configuration files, hardware setup guidelines, testing scripts, and comprehensive data collection capabilities for a wide range of testing scenarios. By advancing beyond traditional security metrics, the soft T-UE endeavors to foster the development of innovative security measures and enhance the ability to anticipate and mitigate potential security breaches in 5G and O-RAN environments. These efforts will culminate in a versatile testing tool capable of evaluating authentication mechanisms, encryption protocols, communication security, logging and monitoring systems, and network component resilience under various attack scenarios.

The structure of this paper unfolds as follows: Section II presents the system architecture and design of the soft T-UE. Section III details the testing methods and associated procedures. Section IV delves into implementation specifics and discusses the results obtained. Finally, the paper concludes by examining the potential impact of the soft T-UE on advancing 5G and O-RAN security testing methodologies.

\section{%Research
System Architecture and Design}
\label{sec:architecture}

\begin{figure}[h]
  \centering
  \includegraphics[width=1\linewidth,trim = 4 4 4 4,clip]{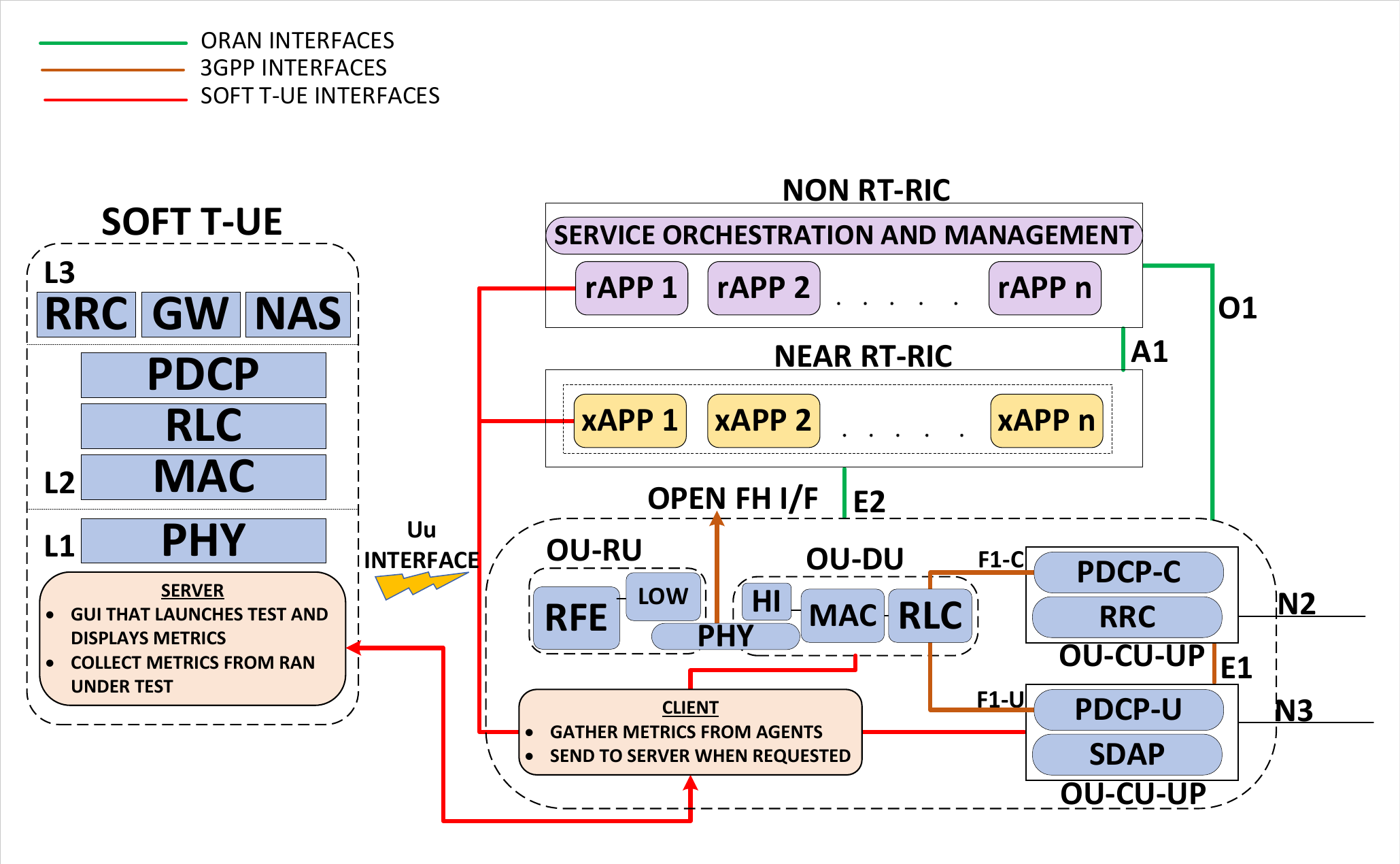}
  \caption{Soft-T UE Architecture}
  \label{fig:arch}
\end{figure}

The architecture of the soft T-UE is engineered to enable rigorous testing of 5G and O-RAN security vulnerabilities without modification to the existing RAN architecture. Built upon srsRAN's User Equipment (UE) structure, it modifies UE functionalities, messages, and signaling mechanisms to conduct precise tests on the RAN components and security procedures under examination. The soft T-UE is designed with modular architecture%ity in mind
, allowing for easy integration of new features and capabilities. The modular approach enables %developers to extend 
the extension of the functionality of the soft T-UE through %adding
new test scenarios, metrics, and analysis tools without disrupting the core architecture. This flexibility is crucial for enabling the rapid integration of new tests to keep pace with the evolving threat landscape of 5G and O-RAN and the rapid development of O-RAN specifications and test requirements. Figure \ref{fig:arch} shows the soft-T UE's architecture from a high-level perspective showing the new interfaces provided, the components used to facilitate data gathering, and the visualization of live metrics.

The soft T-UE connects to the Radio Access Network (RAN) under test by probing the network and sending test vectors that represent different security tests or attacks. The connection establishment protocol in the soft T-UE follows the general 5G/Open RAN attach and connection establishment procedures. This includes initial access procedures, random access, Radio Resource Control (RRC) connection setup, security mode setup, authentication, and NAS procedures. %During initial access, the UE searches for available networks, selects the strongest cell, and initiates the connection. The random access procedure involves sending a preamble and receiving a temporary C-RNTI, followed by RRC connection requests and setup, which involve exchanging UE identity and establishment cause. 
The security mode setup ensures encryption and integrity protection before completing authentication and NAS procedures, such as attach requests and PDU session establishments. The RAN’s responses to these vectors are logged, providing valuable data for both black box and white box RANs. Through soft T-UE, %By
we can introduce a modified version of the regular UE exchanged packets with the RAN during the connection establishment phase or even after this while evaluating the various %new 
default security metrics or even defining new metrics and collecting data during each test for a comprehensive test environment. %, the architecture supports advanced analysis and evaluation of RAN security. 
This data-centric approach allows for a thorough understanding of the RAN's resilience to various types of attacks. The components of the soft T-UE’s architecture include:

%The soft T-UE’s architecture supports a modular design to enhance flexibility and expandability. 

\begin{itemize}
  \item \textbf{UE Functionality:} The soft T-UE functions as a standard UE, with additional capabilities for launching security tests. It supports 4G LTE, 5G NSA, and SA configurations, and can operate in black box, white box, and test configurations. The soft T-UE leverages the existing srsRAN UE architecture, ensuring compatibility with commercial off-the-shelf software radio hardware.
  \item \textbf{Logging and Data Collection:} The soft T-UE architecture features robust logging and data collection mechanisms essential for a comprehensive assessment of the RAN's performance and security. Central to this setup is a client-server model facilitating communication between the RAN under test and the soft T-UE. The client component accepts asynchronous connections from multiple agents concurrently, enabling simultaneous testing of multiple components within each attack scenario. This approach enhances testing scalability and operational efficiency across varied RAN environments. The server component manages incoming data streams, logging real-time UE status, operational parameters, and key performance indicators (KPIs) in JSON format.
  \item \textbf{Graphical User Interface (GUI):} The GUI is an integral component of the soft T-UE architecture. Designed to allow users to easily select and configure tests while providing real-time data visualization, the GUI uses Grafana to display key metrics such as UL/DL bitrate, connection status, attack type, and duration. The architecture includes an interprocess communication (IPC) channel that facilitates real-time data exchange between the soft T-UE and the GUI. JSON-encoded messages are used to update the GUI with the current status and operational metrics of the soft T-UE. This interface ensures that users can monitor and analyze tests effectively, enhancing the usability and accessibility of the soft T-UE. Real-time updates and visual feedback are essential for understanding the test outcomes and making informed decisions based on the results.
  \item \textbf{ Agents in key O-RAN components:} Strategic deployment of agents within O-RAN deployments enhances the functionality and adaptability of the soft T-UE for rigorous security testing. Leveraging the foundational structure of srsRAN's UE framework, the soft T-UE augments UE functionalities, message handling, and signaling mechanisms to conduct precise tests on targeted RAN components. These agents are strategically positioned at critical points within the O-RAN architecture, such as xApps, rApps, or the CU/DU, facilitating the execution of custom %codelets 
  sequence of machine instructions and seamless data collection in response to UE-side signaling and exchanged packets. 
  % This deployment strategy enhances the soft T-UE's adaptability and performance across diverse testing scenarios, making it a versatile tool for comprehensive security evaluations.
\end{itemize}

%\section{From Legacy Cellular Networks  to %the 
%O-RAN} %Architecture}

%\section{O-RAN Resource}
%\label{sec:resources}
%\input{include/resources.tex}

\section{%What 
\textcolor{black}{Testing Methods and Procedures
}}
\label{sec:testing}
\begin{figure}[h]
\centering
\includegraphics[width=0.5\textwidth,trim = 4 4 4 4,clip]{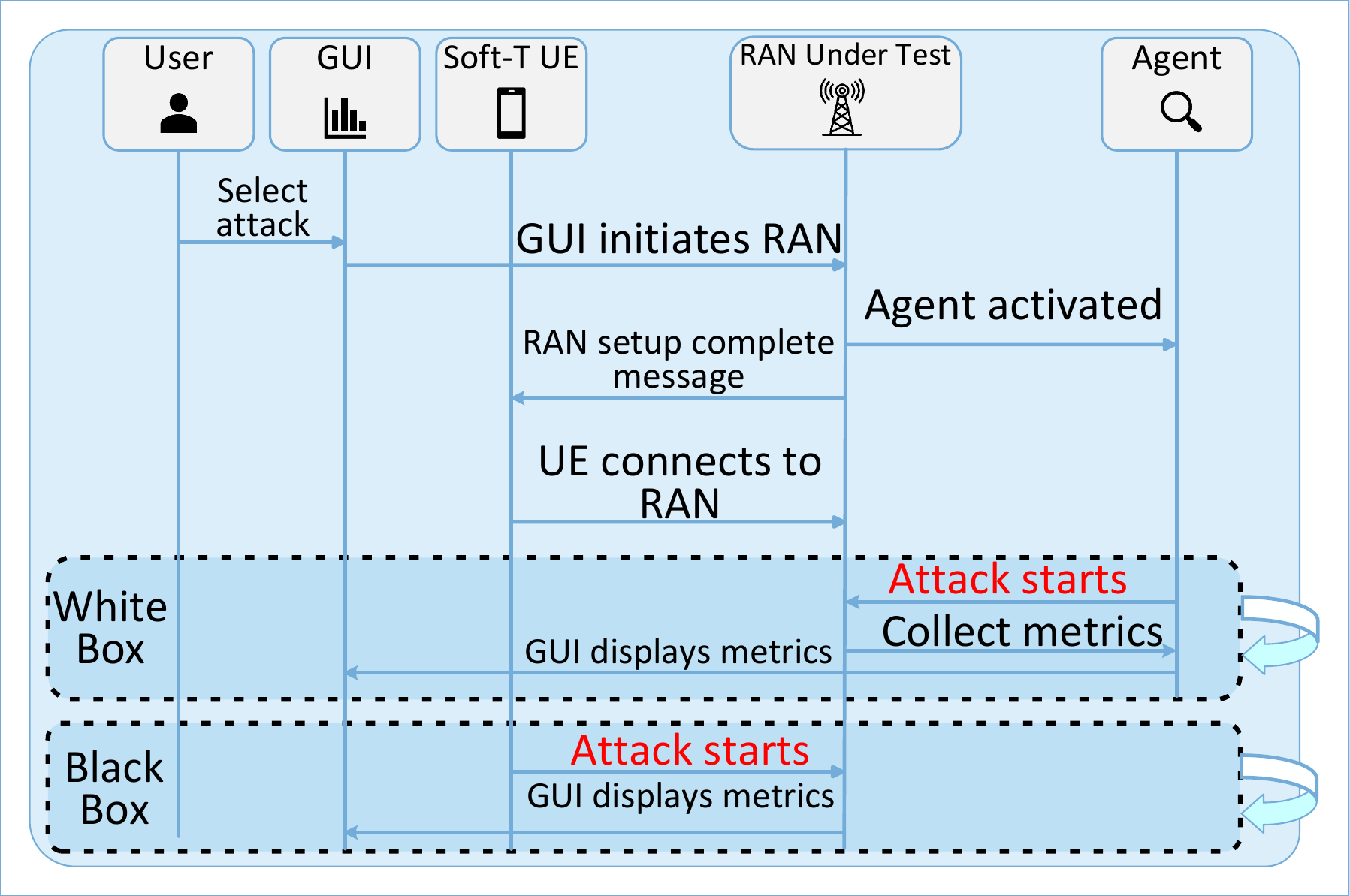}
\caption{General Test Flow}
\label{general test flow}
\end{figure}

The testing methods and procedures supported by the soft T-UE encompass 3GPP Interface testing, O-RAN Alliance test procedures, and custom security test scenarios. These procedures are meticulously designed to evaluate the functionality, conformance, performance, and security of 5G and open RAN components, ensuring a comprehensive and robust testing framework.

\subsection{Test Procedures}

\begin{itemize}
\item \textbf{3GPP Interfaces Testing}: Testing the interfaces specified by 3GPP is crucial for ensuring the security and reliability of 5G networks. These interfaces, such as N1, N2, N3, and others defined by 3GPP, are essential for connecting various network functions and exchanging sensitive data. Extensive vulnerabilities in these interfaces highlight the need for thorough testing \cite{5ginterfaces}. Effective testing procedures aim to verify the integrity, confidentiality, and availability of transmitted data while mitigating potential risks through these interfaces.

\item \textbf{O-RAN Alliance Security Test Specifications}: The soft T-UE incorporates test specifications developed by the O-RAN Alliance’s Working Group 11 (WG11)~\cite{9839628}. These specifications emphasize comprehensive security testing, including vulnerability assessments, penetration testing, and compliance with security requirements. WG11’s frameworks address potential security threats and ensure robust encryption, authentication, and integrity protection mechanisms are in place to safeguard the RAN \cite{oran_security_2024}.

\item \textbf{Custom Security Test Scenarios}: The soft T-UE is equipped to support various configurable attack scenarios, including UE fuzzer, DoS, and spoofing attacks. Users can activate specific parameters to test different security aspects of the RAN, allowing the soft T-UE to adapt to emerging threats and evolving security requirements. This flexibility ensures that the soft T-UE can conduct a wide range of tests, from common security evaluations to custom test procedures designed. %to uncover potential vulnerabilities.

\end{itemize}

\subsection{General Soft T-UE Test Flow}

In the following, we detail the flow of a typical test that can be performed via the soft T-UE tool, including how to employ the GUI or CLI to launch a test, test activating process, and obtain logging results for analysis. Depending on whether white box or black box testing is being conducted, the methodology for implying the attack scenario might change.  %different components perform the attacks. 
The general flow of deploying the security testing use cases using the soft T-UE %attack series of events 
is visualized in Fig.~\ref{general test flow}. The flow can be presented as the following sequence of events:

\begin{enumerate}
  \item \textbf{Configure Test Parameters}: Users begin by configuring the test parameters via the GUI or CLI. The GUI allows for easy selection and configuration of test scenarios, attack types, and specific parameters to be tested, such as the interval between attack instances and the duration of each attack. The CLI offers similar flexibility, enabling users to set parameters through command-line arguments. After the parameters of interest are configured, the test is launched. 
  % \item \textbf{Launch the Test}: Once the configuration is complete, the test is launched by either clicking the ‘Start Test’ button in the GUI or executing the appropriate command in the CLI.
  \item \textbf{Establish Connection}: Upon launching the test, the Soft T-UE initiates the connection establishment procedure with the RAN under test. This involves scanning for available networks, selecting the strongest cell, and completing the standard 5G/Open RAN attach and connection procedures.
  \item \textbf{Execute Test Scenarios}: Once connected, the soft T-UE executes the configured test scenarios. For white box testing, an agent at the component under test performs the attack. For black box testing, the UE itself performs the attack through signaling and packets exchanged with the RAN. %This differentiation allows for a comprehensive evaluation of the RAN's security.
  \item \textbf{Monitor Real-Time Data}: During the test, real-time data is displayed on the GUI or output in the CLI, including UL/DL bitrate, connection status, attack type, duration, and other performance metrics of interest to the specific ongoing test that can be chosen, specifically, at the first step. This real-time feedback allows users to monitor the test progress and performance.
  \item \textbf{Data Collection}: Throughout the test, the soft T-UE collects data on various metrics, such as UL/DL bitrate, connection stability, and RAN responses to attacks. This data is logged in JSON format for easy parsing and analysis. For white box testing, an agent in the white box RAN is activated at the component under test to facilitate data collection and communication with the client residing in the RAN. This ensures accurate and comprehensive data capture, enhancing the reliability of the test results. For black box testing, RAn signaling and responses to the soft T-UE packets are collected. 
  \item \textbf{Data Analysis}: After the test, the collected data is analyzed to evaluate the RAN’s performance and security resilience. The results are documented, and any anomalies or vulnerabilities are identified for further investigation.
\end{enumerate}

\section{{\textcolor{black}{Soft T-UE Use Case Implementation and Results}} %What O-RAN %Can and 
%Cannot Do%\hl{---Limited Capabilities/Features}
}
\label{sec:implementation}
Fuzzing the RRC layer in the UE is a powerful technique to uncover vulnerabilities and ensure robustness in communication protocols. The RRC layer is critical for managing the signaling between the UE and the gNodeB (gNB) in the RAN. This layer handles essential functions such as connection setup, maintenance, reconfiguration, and release. By introducing malformed or unexpected inputs into the RRC messages, fuzzing can identify weaknesses that could be exploited to disrupt network operations, degrade service quality, or compromise security.

When the RRC layer in the UE is subjected to fuzzing, several RAN functions can be significantly impacted. The gNB, which manages RRC connections with the UE, may encounter malformed %or unexpected 
RRC connection requests, reconfiguration messages, and connection releases. Such disruptions can cause gNB malfunctions or crashes, compromising the %stability and 
reliability of the network. %Furthermore, the gNB manages radio bearers, which are crucial for data and control plane communications. 
Fuzzed RRC messages can disrupt the establishment, modification, and release of %these 
radio bearers managed by the gNB, causing interruptions in data transmission and signaling. Additionally, handover procedures %, which ensure seamless transitions of UEs 
between cells can be disrupted, resulting in dropped connections and degraded service quality. Security management is another critical area where the gNB's role is vital, and fuzzing RRC messages related to security can potentially compromise the integrity and confidentiality of communications~\cite{10154283}.

The impacts of RRC fuzzing extend beyond the RAN to core network components. The Access and Mobility Management Function (AMF), which handles UE registration, deregistration, and mobility management, can experience abnormal behavior due to malformed RRC messages. This can lead to issues with registration, mobility, and session management, causing service disruptions \cite{10449510}. The AMF is also responsible for authenticating UEs and managing security contexts. Fuzzed RRC messages can interfere with these procedures, potentially leading to security breaches targeting the AMF functionalities and other core network functions, such as the Session Management Function (SMF), Unified Data Management (UDM), and Policy Control Function (PCF). 
% The Session Management Function (SMF), which manages PDU session establishment, modification, and release, can also be affected, resulting in incorrect or failed session management procedures. The Unified Data Management (UDM) component, which stores and manages subscriber data, might face inconsistencies or errors in data handling due to disrupted RRC signaling. Similarly, the Policy Control Function (PCF), responsible for enforcing policies and managing Quality of Service (QoS) parameters, can experience issues, leading to suboptimal service delivery.

We specifically targeted the initial UE message registration request for fuzzing to evaluate the robustness of the RRC layer. This focused approach allowed us to assess how the gNB handles unexpected inputs specifically during the registration process, examining its impact on critical functions. During the test, we are utilizing %hardware and software setup provided 
a controlled environment for fuzzing, ensuring that the tests only affect the initial UE message registration request to accurately capture the standalone effect of misconfiguration RRC on the performance of the RAN. This controlled approach ensured systematic testing of the RRC layer under varied conditions, facilitating the identification of vulnerabilities and assessment of network robustness which aligns with recent literature attempting to formalize the fuzzing process \cite{10444528}.
%live network operations.
Extensive logging and monitoring were employed to capture and analyze the behavior of both the RAN and core network components during the fuzzing process. %This analysis was crucial in identifying vulnerabilities and potential security risks, allowing for necessary mitigation measures to enhance network resilience and security.

\subsection{Implementation}

\begin{figure}[ht]
  \centering
  \includegraphics[width=0.5\textwidth,trim = 4 4 4 4,clip]{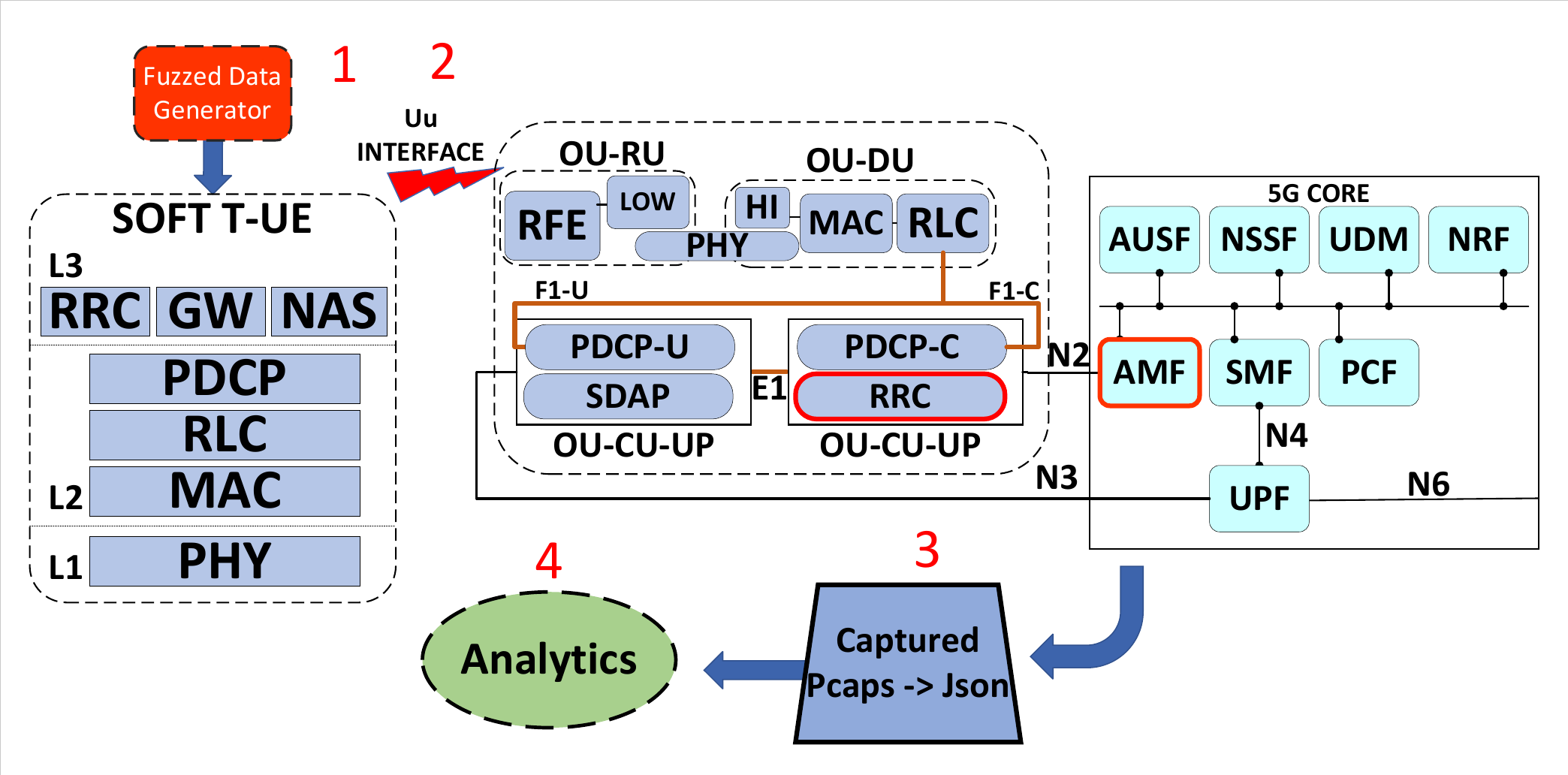}
  \caption{Fuzzing attack diagram}
  \label{fig:attack}
\end{figure}

For the implementation of RRC fuzzing, we utilized SRSRAN Project version 24.4 for the gNB, leveraging an Ettus B210 USRP as the hardware platform. For the UE, we employed the soft-T UE, which was modified from SRS UE version 23.11, also using an Ettus B210 USRP to ensure compatibility and optimal performance. The core network was Open5GS release 17. All software was running on COTS equipment running the latest LTS release of Ubuntu Linux paired with a low-latency kernel. This setup allowed us to simulate and fuzz the RRC messages effectively between the UE and the RAN. The implementation can be seen in Fig.~\ref{fig:attack} which involved the following procedures executing in order:

\begin{enumerate}
  \item A fuzzer on the soft-T UE specifically targets and manipulates the initial UE message registration request, ensuring comprehensive testing of the RRC layer's handling of registration procedures.
  \item The fuzzed registration messages are transmitted to the RAN through the Uu interface, allowing for the evaluation of the gNB's response and behavior under various fuzzing conditions.
  \item  Network traffic in the form of PCAP files is captured during the fuzzing process. These PCAP files are then converted to JSON format to facilitate seamless integration and analysis within the GUI.
  \item Critical metrics such as connection success rates and packet error rates are plotted on the GUI. This provides users with a real-time, visual representation of the system's performance and robustness under various test conditions.
\end{enumerate}

\subsection{Results Analysis}

\begin{figure}[h]
  \centering
  \includegraphics[width=0.45\textwidth]{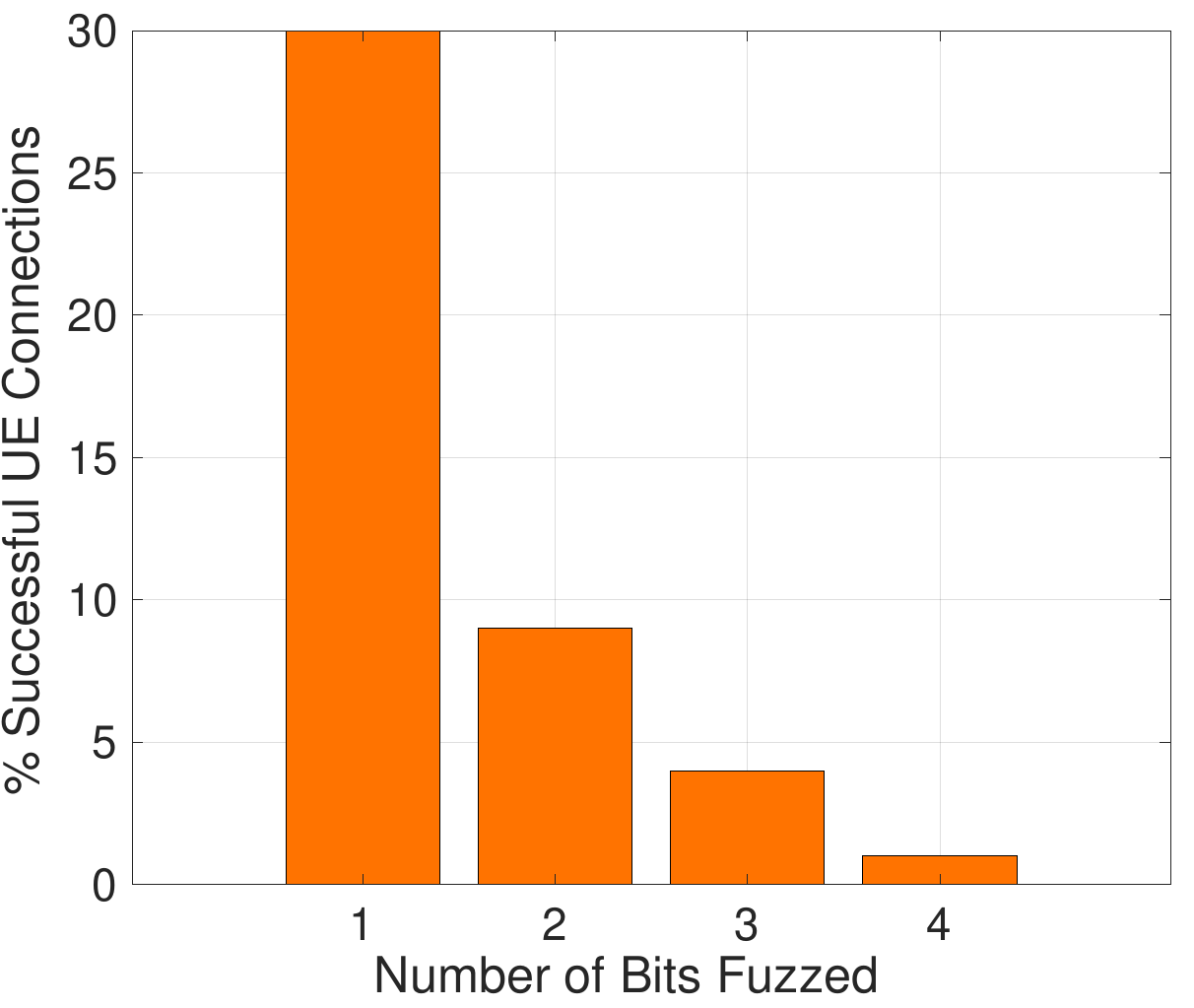}
  \caption{Effect of fuzzing}
  \label{fig:Effect of fuzzing}
\end{figure}

Fuzzing the RRC layer in the UE reveals how critical the robustness and security of signaling protocols are for the overall stability of the network. We performed fuzzing tests on the initial UE message registration request, focusing on the 26 data bytes (208 bits) that constitute the data portion of the packet. To ensure a thorough analysis, we conducted 100 tests randomly altering a certain number of bits each time. Automating the UE fuzzing process was achieved using a custom controller that initiates the Soft T-UE and the RAN, followed by a modified function for the SDU buffer. This function targets a byte within the SDU buffer data, such as RRC registration request and other RRC messages, and fuzzes one of the 8 bits in that byte to introduce variability and detect potential vulnerabilities. During each test the RAN's response was monitored for anomalies, such as unexpected acceptance of malformed requests, system crashes, or other abnormal behaviors. Our objective was to identify vulnerabilities in the registration process that could be exploited by malicious actors. We measured the impact on registration success rates. Our analysis revealed that only a few bits altered had a disproportionately high impact on successful UE connections which can be seen in Fig.~\ref{fig:Effect of fuzzing}. The RRC Setup Complete message, which is 26 bytes in length and contains the RRC Registration Request which is encrypted, underwent bitwise fuzzing as shown in Fig. 5. Each bit within every byte was altered individually, revealing the vulnerability shown from 1 to 100, representing the chance that the UE will still create a PDU session with the corrupted data with 100 being the most likely to still work. This analysis highlights the protocol's significant vulnerability to fuzzing, with a high likelihood of failure for most bits. Furthermore, this result suggests that a targeted fuzzing approach could induce failures more frequently. This comprehensive approach underscores the importance of detailed and systematic testing in uncovering and mitigating potential security weaknesses in essential network operations.

\begin{figure}[h]
  \centering
  \includegraphics[width=0.48\textwidth]{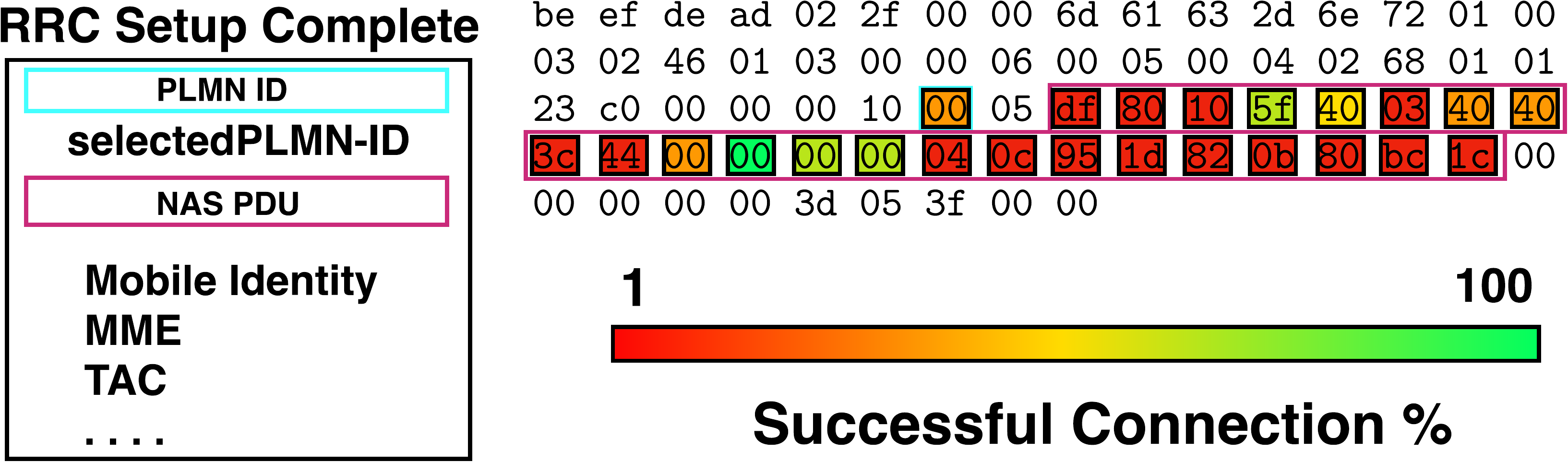}
  \caption{Vulnerability of bits fuzzed}
  \label{fig:Effect of fuzzing2}
\end{figure}

%\section{Testbed Resources}
%\label{sec:testbed}
%\input{include/testbeds}

% \section{O-RAN Evolution: To Enable the Requested Features} %Architecture}
% \label{sec:contribution}
% \input{include/contribution.tex}

% \section{Research Directions to O-RAN}
% \label{sec:case}
% \input{include/directions.tex}

\section{Conclusions}
\label{sec:conclusions}
In conclusion, the rapid advancements in networks necessitate robust and flexible security testing mechanisms. Traditional testing approaches, hindered by high costs and proprietary constraints, are insufficient for the dynamic and complex threat landscape faced by these modern networks. The %introduction of the 
proposed soft T-UE addresses these challenges by offering a versatile, open-source platform for comprehensive security evaluations. The soft T-UE empowers the 
 %researchers and practitioners with the tools needed to conduct thorough 
assessment of authentication mechanisms, encryption protocols, and the resilience of network components under diverse attack scenarios. %By advancing beyond traditional security metrics, t
The soft T-UE fosters innovation in security measures, enhancing the ability to anticipate and mitigate potential breaches in 5G and O-RAN environments. This paper has outlined the system architecture, detailed the testing methods, and provided implementation specifics, demonstrating the potential of the soft T-UE to significantly improve security testing practices in the rapidly evolving landscape of wireless communications.

\section*{Acknowledgement}
This material is based upon work supported by the National Telecommunications and Information Administration (NTIA) under Award No. 28-60-IF012

\balance

\bibliographystyle{IEEEtran}
\bibliography{main}

\section*{Biographies}
\footnotesize
\vspace{0.2cm}
\noindent
\textbf{Joshua Moore} (jjm702@msstate.edu)
is a PhD student in the Department of Electrical and Computer Engineering at Mississippi State University, Starkville, MS, USA. His research interests include O-RAN, 5G/next-G communications, and RAN Management and Orchestration.

\vspace{0.2cm}
\noindent
\textbf{Aly Sabri Abdalla} (asa298@msstate.edu)
is an Assistant Research Professor in the Department of Electrical and Computer Engineering at Mississippi State University, Starkville, MS, USA. His research interests are on wireless communication and networking, software radio, spectrum sharing, wireless testbeds and testing, and wireless security with application to mission-critical communications, open radio access network (O-RAN), unmanned aerial vehicles (UAVs), and reconfigurable intelligent surfaces (RISs).

\vspace{0.2cm}
\noindent
\textbf{Charles Ueltschey}
(cmu32@msstate.edu) is an undergraduate student in computer science at Mississippi State University, Starkville MS, USA. His research interests are Wireless communications, O-RAN security, and Machine Learning.

\vspace{0.2cm}
\noindent
\textbf{Vuk Marojevic} (vuk.marojevic@msstate.edu) is an associate professor in electrical and computer engineering at Mississippi State University, Starkville, MS, USA. His research interests include resource management, vehicle-to-everything communications and wireless security with application to cellular communications, mission-critical networks, and unmanned aircraft systems.

\end{document}